\documentstyle[preprint,aps]{revtex}

\begin{document}

\preprint{INP 1623/PH, DOE/ER/40427-07-N93, TPR--93--12}

\draft
\tighten

\title{$N_c$-counting rules and the axial vector coupling constant\\
       of the constituent quark}

\author{Wojciech Broniowski \cite{ifj}\cite{humboldt}}

\address{Institute of Theoretical Physics, University of Regensburg,
         D-8400 Regensburg, Germany}

\author{Matthias Lutz \cite{ml}}
\address{Department of Physics and Astronomy, University of Washington,
         WA 98195, USA}

\author{Andreas Steiner}

\address{Institute of Theoretical Physics, University of Regensburg,
         D-8400 Regensburg, Germany}

\date{April-May, 1993}
\maketitle

\begin{abstract}
We analyze the axial vector coupling of the constituent quark
in the large-$N_c$ limit and point out mechanisms
which yield $1 - g_A \sim N_c^0$.
\end{abstract}


\narrowtext

The concept of {\em constituent} quarks has been useful
in various models of hadronic structure, leading to a quite successful
description of low-energy spectroscopy.
It is generally assumed that these constituent quarks are Dirac particles,
{\em i.e.} have no anomalous magnetic moment, and their axial vector
coupling constant, $g_A$, equals~$1$.
This is puzzling in view of the fact that
{\em constituency} results from complicated, nonperturbative mechanisms,
which {\em inter alia} generate the large constituent mass of
roughly one third the nucleon mass.

In recent {\em Letters} Weinberg
argued why constituent quarks should behave as Dirac particles
\cite{Weinberg:gA,Weinberg:nc}. In Ref. \cite{Weinberg:nc}, using
the Adler-Weisberger sum rule \cite{Adler,Weisberger} and
large-$N_c$ (number of colors) arguments, he gives an estimate for
the quantity $1-g_A^2$, which, with the low-energy
mechanisms taken into account, is of order $1/N_c$.

In the first part of this {\em Letter} we point out a mechanism which
produces a contribution to $1-g_A$ at the level $N_c^0$:
{\em the $\pi$-$A_1$ mixing mechanism}.
This mechanism results from very fundamental facts:
1)~the presence of vector mesons, which in the
large-$N_c$ limit are treated as $q \overline{q}$ bound states, and
2)~spontaneous breaking of the chiral symmetry.
We present an estimate of $1-g_A$ in models having the above
mentioned features:
the $\sigma$-model with vector mesons \cite{benlee,BB85},
and the Nambu--Jona-Lasinio (NJL) model \cite{Vogl1,Vogl2}.

In the second part of this {\em Letter} we follow Weinberg and
use the Adler-Weisberger sum rule approach. Again, we
demonstrate that there are contributions to the
sum rule which yield $1-g_A^2 \sim N_c^0$. These effects are
dominant at high-energies, and we model them with
the reggeized $\rho$-meson
exchange in the t-channel. We estimate the size of these
leading-$N_c$ effects, and find that for $N_c=3$ they are of similar
magnitude as the $1/N_c$ corrections considered by Weinberg
\cite{Weinberg:nc} and other authors \cite{Peris,Dicus}.

Let us start with a brief reminder of the basic
large-$N_c$ counting rules
\cite{thoft:nc,witten:nc,banerjee:nc} which we need below:
meson masses scale as $N_c^0$, the gluon-quark and meson-quark
coupling constants scale as $1/\sqrt{N_c}$,
the $n$-meson vertex scales as $N_c^{1-n/2}$, the pion decay
constant, $F_\pi$, scales as $\sqrt{N_c}$, hadronic sizes scale
as $N_c^0$.
One of the crucial points behind the large-$N_c$ philosophy is
that all (infinitely many) mesons that can be formed
of a $q \overline{q}$ pair should be included
in low-energy phenomenology, most importantly
the low-lying vector mesons.
Hence, in the large-$N_c$ approach, vector mesons are treated as
$q \overline{q}$ bound states.
This is in contrast to the treatment of vector
mesons as multipion resonances. We will come back to this issue
at the end of this {\em Letter}.

Let us now show how $1-g_A$ obtains contributions of the order
$N_c^0$. A  Goldberger-Treiman \cite{ward:end}
relation links $g_A$ to
the pseudovector pion quark coupling constant, $g_{\pi q}^A $, the
pion decay constant, $F_\pi $, and the constituent quark mass, $m$:
\begin{equation}\label{gener}
1-g_A = g_{\pi q}^A F_\pi / m \; .
\end{equation}
A nonvanishing $g_{\pi q}^A $ is generated by
the $\pi$-$A_1$ mixing mechanism. This is easily illustrated in the
$SU(2) \times SU(2)$ $\sigma$-model with $\rho$ and $A_1$ vector mesons
described in Ref.~\cite{benlee,BB85}. The lagrangian, to be used at the tree
level, has the form
\begin{eqnarray}
 {\cal L} &=& \bar{\psi} \left [
\dot{\imath} \overlay{\slash}{\partial}
   + g_{\pi} \left ( \sigma+\dot{\imath}
{\gamma_{5}} \mbox{\boldmath $\tau$} \cdot
\mbox{\boldmath $\pi$} \right )
   + \mbox{$1\over 2$} g_{\rho} \mbox{\boldmath $\tau$} \cdot
     \left ( \overlay{\slash}{\mbox{\boldmath $\rho$}} +
             \gamma_5 \overlay{\slash}{\mbox{\boldmath $A$}} \right )
     \right ] \psi \nonumber \\
    &+&  \mbox{$\beta \over 2$} \left ( (D^{\mu} \sigma)^{2}
       + (D^{\mu} \mbox{\boldmath $\pi$})^{2} \right )
       - U \left ( \sigma, \mbox{\boldmath $\pi$} \right ) \nonumber \\
    &-&  \mbox{$1\over 4$} (\rho^{\mu \nu})^2
     -   \mbox{$1\over 4$} (A^{\mu \nu})^2
     +  \mbox{$1\over 2$} m_{\rho}^2 \left (
         (\rho^{\mu})^2 + (A^{\mu})^2 \right ) ,
    \label{eq:GML}
\end{eqnarray}
where $\psi$ is the constituent quark field, $\sigma$, $\pi$,
$A$ and $\rho$ are the meson fields, and the constant $\beta$
is explained below.
The potential $U$ leads to spontaneous chiral symmetry breaking, and
the quark acquires its ``constituent'' mass,
\mbox{$m = - g_{\pi} \sigma_{vac} = g_{\pi} F_{\pi}$} (our convention is
$F_\pi = 93$ MeV). The covariant derivative of the pion field has the form
\begin{eqnarray}
%   D^{\mu} \sigma &=& \partial{}^{\mu} \sigma
%   + g_{\rho} \mbox{\boldmath $A$}^{\mu} \cdot \mbox{\boldmath $\pi$} , \\
   D^{\mu} \mbox{\boldmath $\pi$} &=&
          \partial{}^{\mu} \mbox{\boldmath $\pi$}
   + g_{\rho} \left (
       \mbox{\boldmath $\rho$}^{\mu} \times \mbox{\boldmath $\pi$}
     - \mbox{\boldmath $A$}^{\mu} \sigma \right ) \nonumber \\ &=&
          \partial{}^{\mu} \mbox{\boldmath $\pi$}
   + g_{\rho} F_{\pi} \mbox{\boldmath $A$}^{\mu} + ...
\label{eq:cov}
\end{eqnarray}
where in the last equality we have expanded the $\sigma$ field around
its vacuum value in order to visualize the $\pi-A_1$ mixing.
Appropriate expressions for $D^{\mu} \sigma$, field-strength
tensors $\rho^{\mu \nu}$,
$A^{\mu \nu}$, and the potential $U$, which are not relevant here,
can be found in Ref. \cite{benlee,BB85}.
Diagonalization of the $\pi$ and $A_1$ propagation
is achieved by introducing the physical $A_1$ field
\begin{equation}
\mbox{\boldmath $A$}^{\mu}_{\text{ph}}
 = \mbox{\boldmath $A$}^{\mu}
+ \frac{g_{\rho} F_{\pi}}{m_{\rho}^2}
  ( \partial{}^{\mu} \mbox{\boldmath $\pi$}
   + g_{\rho} \mbox{\boldmath $\rho$}^{\mu}
 \times \mbox{\boldmath $\pi$}).
\label{eq:aphys}
\end{equation}
In order to ensure the proper wave-function normalization of the pion field,
the constant $\beta$ in Eq. (\ref{eq:GML}) has to be equal to
\mbox{$\beta =
m_{\rho}^2 / (m_{\rho}^2 - g_{\rho}^2 F_{\pi}^2)$} \cite{benlee}.
The KSFR relation requires $\beta=2$, and leads to the
Weinberg relation $m_{A^{ph}} = \sqrt{\beta} m_{\rho} = \sqrt{2} m_{\rho}$.
Now, with the physical degrees of freedom,
the quark-pion coupling term in the lagrangian has the form
\begin{equation}
 {\cal L}_{\text{q}\pi} = \bar{\psi} \left [
   g_{\pi} \dot{\imath}
\gamma_{5} \mbox{\boldmath $\tau$} \cdot
\mbox{\boldmath $\pi$}
   - \mbox{$1\over 2$} \frac{g_{\rho}^2 F_{\pi}}{m_{\rho}^2}
    \mbox{\boldmath $\tau$} \cdot
     \gamma_5 \gamma_{\mu} \partial{}^{\mu} \mbox{\boldmath $\pi$}
    \right ] \psi .
\label{eq:qm}
\end{equation}
The second term in the above equation is generated by the $\pi$-$A_1$
mixing mechanism. Identifying the pseudovector coupling constant
$g_{\pi q}^A =m g_\rho ^2 F_\pi /{m_\rho }^2 $ of Eq. (\ref{gener}) leads
to the following result:
\begin{equation}
1-g_A = \frac{g_{\rho}^2 F_{\pi}^2}{m_{\rho}^2} \sim N_c^0 .
\label{eq:ward}
\end{equation}
According to the $N_c$-counting rules, expressions (\ref{gener}) and
(\ref{eq:ward}) scale as
$N_c^0$, hence the $\pi$-$A_1$ mixing mechanism
shifts the value of $g_A$ from $1$ {\em at order $0$ in $N_c$}.
With our expressions for $\beta$ and $m_{A^{ph}}$ we may also write
\begin{equation}
g_A = \frac{m_{\rho}^2}{m_{A^{ph}}^2},
\label{eq:ward1}
\end{equation}
Using empirical numbers we would obtain
$g_A \simeq 0.4$, a very small value. This may be due to
the treatment of vector fields as static fields.
The NJL model, for example,
treats mesons dynamically, and leads to Eq. (\ref{eq:ward1}) \cite{Klimt}
with the meson masses
replaced by ``off-shell masses'' evaluated at $q^2 = 0$. These are much
different from the physical masses \cite{Jaminon}.
As a result, the NJL model yields the
following estimate for the axial
vector coupling constant of the quark: $g_A \simeq 0.7 - 0.8$.
Explicit formulas in this model confirm the basic result that
$1-g_A \simeq N_c^0$
(see Ref. \cite{Vogl1,Vogl2} for details).

In the NJL model, the quark acquires an anomalous magnetic moment only
if the 't Hooft $U(1)$-breaking interaction is introduced. However,
the numerical value is very small \cite{Vogl1,Vogl2}.

In the remaining part of this {\em Letter} we follow
Weinberg's approach \cite{Weinberg:nc}, and focus on the Adler-Weisberger
sum rule for the constituent quark \cite{rem:confinement}:
\begin{equation}
1 - g_A^2 = \frac{2 F_{\pi}^2}{\pi} \int_0^{\infty}
\frac{d\omega}{\omega} \left (
\sigma_{-}(\omega) - \sigma_{+}(\omega) \right ) ,
\label{eq:AWSR}
\end{equation}
where $\sigma_{\pm}(\omega)$ is the total cross section for scattering
of $\pi^{\pm}$ of energy $\omega$ on an {\em up} quark at rest.
At low energies elastic scattering is dominant \cite{Weinberg:nc}, and
one can use the effective low-energy lagrangian with pions and quarks
\cite{Weinberg:physica,ManoharG,BirBan84} to calculate the cross sections in
(\ref{eq:AWSR}). An example of such a process is illustrated in
Fig.~\ref{elreg}(a). The top diagram is in the hadronic representation,
the middle diagram shows a typical QCD contribution to this process, and
the bottom diagram is the corresponding index-line representation
of 't Hooft \cite{thoft:nc}. The cross
section for the process in Fig. \ref{elreg}(a)
is of the order $1/N_c^2$, and upon multiplying by the factor
$F_{\pi}^2$ in Eq. (\ref{eq:AWSR}) gives a contribution to
$1 - g_A^2$ of the order $1/N_c$. Dicus {\em et al.} \cite{Dicus}
considered also the inelastic process of quark-antiquark production, which
enters at the same $N_c$-level, and found it to be numerically small.
The numerical estimate of Ref. \cite{Weinberg:nc} gives
$1 - g_A^2 = 0.17 \sim 1/N_c$ \cite{rem:infty}.

It is, however, possible to find processes contributing to
the sum rule (\ref{eq:AWSR}) at the level $N_c^0$. An example is
shown in Fig. \ref{elreg}(b), which represents the imaginary part
of the forward pion-quark scattering amplitude. Through the
optical theorem this gives the relevant cross sections for
Eq. (\ref{eq:AWSR}). Clearly, this cross section is of the order
$1/N_c$, and the corresponding contribution to $1-g_A^2$ is of the
order $N_c^0$.

The diagrams in Fig. \ref{elreg}(b) are purposely drawn in such a way
that the object exchanged between the quark and the pion looks as a meson,
or, more appropriately, as a {\em reggeon}.
Recalling the success of Regge phenomenology in describing
the large-energy charge-exchange
pion-nucleon scattering \cite{Hoeler,Collins}, it seems
adequate to
transplant these concepts to our pion-quark system. It has
already been assumed in Eq. (\ref{eq:GML}) or (\ref{eq:AWSR})
that the quark can be treated as a free particle, and it is only
different from the nucleon in its mass and coupling parameters.
At high energies Reggeology dominates the quantity
$\sigma^{(-)} = \sigma_{+} - \sigma_{-}$,
and the total cross section can be obtained from the
imaginary part of the forward
scattering amplitude of the $\rho$-reggeon exchange.

Of course, the pion-quark scattering cross section cannot be obtained
from experiment. However, we may use the concept of {\em universality
of the reggeon coupling} to relate this cross section to the
pion-nucleon cross section. The basic result, found for a large
variety of hadronic reactions \cite{russians},
is that the total reggeon exchange
cross section for scattering of particle $a$ on particle $b$ behaves as
\begin{equation}
\sigma_{ab}(s) = \frac{\pi}{\overline{s}} \,g_a(0) \,g_b(0)
\left ( \frac{s} {s_0^{ab}} \right )^{\alpha_0-1},
\label{eq:rus}
\end{equation}
where $g_a(0)$ and $g_b(0)$ are the coupling constants of the reggeon
to the target and the projectile at $t=0$, $\alpha_0$ is the Regge intercept,
and $\overline{s}$ is an overall constant
independent of $a$ and $b$.
The quantity $s_0^{ab}$ has the form
\begin{equation}
s_0^{ab} = \frac{\Lambda^2}{\overline{x}_a \overline{x}_b},
\label{eq:s}
\end{equation}
where $1/ \overline{x}_a$ and  $1/ \overline{x}_b$ are the average
number of quarks in the target and projectile, and $\Lambda$
is some energy scale, independent of $a$ and $b$.
In Ref.~\cite{russians}
it has been found that ratios of total cross sections for
many hadronic processes at high energies are
remarkably well reproduced
if one assumes Eq. (\ref{eq:s}), and {\em universality} of the coupling
constants $g_a$ and $g_b$. In particular, the $\rho$-reggeon
couples universally to isospin.
In our case, it means
that the $\rho$-reggeon--nucleon, $\rho$-reggeon--quark, and
the $\rho$-reggeon--pion
coupling constants are equal, and we denote them by $g(0)$.
For large values of the pion lab energy
$\omega$ Eq. (\ref{eq:rus}-\ref{eq:s})
give the following expression for the pion-quark scattering:
\begin{equation}
\sigma^{(-)}(\omega) = \frac{\pi}{\overline{s}}\,
g^2(0) \left ( \frac{m \,\omega}{\Lambda^2} \right )^{\alpha_0 - 1} ,
\label{eq:sigmin}
\end{equation}
where we have used $1/ \overline{x}_q = 1$ and
$1/ \overline{x}_\pi = 2$.
Substituting Eq. (\ref{eq:sigmin}) into Eq. (\ref{eq:AWSR}), and integrating
from energy $\omega_0$ to infinity, we obtain
\begin{equation}
(1-g_A^2)_{\text{Regge}} =
\frac{2 F_\pi^2}{\pi (1-\alpha_0)} \, \sigma^{(-)}(\omega_0).
\label{eq:final}
\end{equation}

We have already argued on the basis of
Fig. \ref{elreg}(b) that $\sigma^{(-)} \sim 1/N_c$, which
leads to $(1-g_A^2)_{\text{Regge}} \sim N_c^0$.
The same conclusion follows from Eq. (\ref{eq:sigmin}),
since the parameters $\overline{s}$, and $\Lambda$
scale as $N_c^0$. These parameters may depend on relevant
{\em sizes} in the reaction, which scale as $N_c^0$. All dependencies
on $N_c$ through masses or $\overline{x}$ have been taken into account
explicitly in Eq. (\ref{eq:rus}-\ref{eq:s}) \cite{russians}.
The coupling constant $g(0)$ scales as a generic quark-meson
coupling constant, $g(0) \sim 1/\sqrt{N_c}$.
The quantity $(1-\alpha_0)$ also scales as $N_c^0$,
because for a linear trajectory this quantity is directly related to the
string tension: $\alpha' = (1-\alpha_0)/m_{\rho}^2$, which is independent
of $N_c$. This result follows from
lattice QCD in the strong coupling limit \cite{lattice:string}.
It is also consistent with the scaling rules for meson masses and radii
\cite{thoft:nc,witten:nc,banerjee:nc}.

To obtain
a numerical estimate, we assume that the large-$N_c$ value of $\alpha_0$
does not differ much from
its $N_c=3$ value: $\alpha_0 \simeq 0.5$.
With the use of arguments of Ref. \cite{russians} we may
obtain the numerical contribution to Eq. (\ref{eq:AWSR}) in
a model-independent fashion. This is because the ratio of the quark-pion
to the nucleon-pion cross sections is
\begin{equation}
\frac{\sigma^{(-)}(\omega)}{\sigma^{(-)}_{\pi N}(\omega)} =
\left ( \frac{ m \,\overline{x}_{q}}{M_N \, \overline{x}_{N}}
\right )^{\alpha_0 - 1} \simeq 1 ,
\label{eq:ratio}
\end{equation}
where the last equality follows from the fact that (for $N_c = 3$) the nucleon
mass $M_N \simeq 3 m$, and $\overline{x}_{q} / \overline{x}_{N} = 3$.
To obtain our estimate we proceed as follows. We choose
$\omega_0 = m_\rho$. Then, we fit high-energy pion-nucleon data to the
reggeon-exchange formula and extrapolate $\sigma^{(-)}_{\pi N}$ to the
energy $\omega_0$.
Equations (\ref{eq:final}-\ref{eq:ratio})
yield $(1-g_A^2)_{\text{Regge}} \simeq 0.2$.

At energies below $\omega_0$ elastic processes such as those
of Ref. \cite{Weinberg:nc} should dominate. However, the integration
limits for these processes in Eq. (\ref{eq:AWSR})
should be from $0$ to $\omega_0 \simeq m_{\rho}$, which
leads to a lower estimate of this contribution than the value
obtained in Ref. \cite{Weinberg:nc}, where the integration is
performed up to infinity. We get $(1-g_A^2)_{\text{elastic}} \simeq 0.06$,
a three times lower value, which results from the fact that the integrand
of Ref. \cite{Weinberg:nc} has a long high-energy tail $\sim 1/\omega^2$.

The total estimate obtained from the Adler-Weisberger sum rule may therefore
be written as
\begin{equation}
1-g_A^2 \simeq 0.2 + 0.06 \frac{3}{N_c} .
\label{eq:estimate}
\end{equation}
The precise value of the both terms above depends on
specific values of parameters, and our estimate
should be treated as an
``estimate within factors of~2'' only.
In our estimate, the change of the value of $\omega_0$ to $2 m_{\rho}$ yields
$1-g_A^2 \simeq 0.14 + 0.10 \frac{3}{N_c}$, and to
$3 m_{\rho}$ yields
$1-g_A^2 \simeq 0.12 + 0.12 \frac{3}{N_c}$, hence the total contribution is
insensitive to the exact choice of $\omega_0$.
The value of $g_A$ resulting from the above numbers,
$g_A \simeq 0.86$, is somewhat larger than the value obtained in the NJL
model from the $\pi$-$A_1$ mixing mechanism.

In conclusion, we would like to return to the question whether
vector mesons should be treated as explicit degrees of freedom.
Both our derivations
of the $N_c^0$ contributions to $1-g_A$, the direct one, involving the
$\pi$-$A_1$ mixing mechanism, and the derivation via the Adler-Weisberger
sum rule, rely on the presence of vector meson degrees of freedom.
Furthermore, our assignment of the $N_c$-counting rules implicitly
assume that
vector mesons are pure $q \overline{q}$ bound states. Let us
concentrate on the $\rho$ meson.
The $N_c$-counting rules would be different if we treated the
$\rho$ meson as a $\pi$-$\pi$ bound state. The $\pi$-$\pi$
component of the $\rho$ is subleading in $N_c$, according to the
usual 't Hooft--Witten counting rules ({\em e.g.} the appropriate
index-line diagram for the propagator would involve at
least one index line hole).
Accordingly, the coupling
of a multipion component of a vector meson to the quark is
$N_c$-suppressed compared to the coupling of its pure
$q \overline{q}$ component. The fundamental
question which arises is:
{\em What is the composition of a vector meson at $N_c=3$?}
If, as assumed by the large-$N_c$ models such as the NJL model,
the vector meson is predominantly a $q \overline{q}$ bound state
even at $N_c=3$, then
$N_c$-counting should be useful for vector meson effects in physical world.
On the other hand, if at $N_c=3$ a vector meson is mostly a multipion
resonance, then formally subleading effects in $N_c$ will numerically
dominate. In the extreme case where a vector meson has no pure
$q \overline{q}$ component at all, its couplings are
suppressed in $N_c$, and we recover the
Weinberg result $1-g_A \sim 1/N_c$.
In our numerical estimates in this paper we assumed
that the physical values of
parameters associated with vector mesons correspond
to the leading-$N_c$ terms.
The above mentioned ambiguity may shift the strength between
subsequent terms in expansions such as Eq. (\ref{eq:estimate}).
Note, however, that the total value at $N_c=3$,
coming from a given physical effect, would not change.

A practical feature of the $N_c$-counting
is that it may be used to eliminate certain processes, and
this fact is not affected by the above-discussed problems. For instance,
in our sum rule calculation
at low energies multipion production is suppressed relative to the
included elastic processes, and at high energies a multi-reggeon
exchange is suppressed compared to a single reggeon exchange.

After this work has been completed a preprint by Peris and de Rafael
has been released \cite{PerisRafael}, reporting similar results
obtained from the NJL model. In particular, a formula
similar to Eq. (\ref{eq:ward1}) is given.

We thank Sergei Kulagin and
Klaus Steininger for
useful discussions and Wolfram Weise for many helpful remarks.
We are particularly grateful for
Kolya Nikolaev for pointing out Ref. \cite{russians}
and for helpful comments.
This work has been supported in part by BMFT grant 06 OR 735,
and the Polish State Committee for Scientific Research
grants 2.0204.91.01 and 2.0091.91.01.

\begin{figure}
\caption{(a) A sample low-energy contribution
 to the Adler-Weisberger sum rule:
 elastic pion-quark scattering
\protect{\cite{Weinberg:nc}}.
 Hadronic diagram (top), a typical QCD diagram
 contributing to this process (middle), and the 't Hooft index line
 representation of this diagram (bottom). The cross section scales as
 \mbox{$1/N_c^2$}.
(b) A high-energy contribution,
 obtained via optical theorem from the
 imaginary part of the forward $\pi$-quark scattering amplitude,
 described via reggeized $\rho$ exchange.
 The cross section scales as
 \mbox{$1/N_c$}.
}
\label{elreg}
\end{figure}


\begin{references}
\bibitem{ward:end}
The Goldberger-Treiman relations follow from  PCAC
and from the Ward identity:
\[
q^{\mu }\Lambda_{\mu }^a (p',p)
=S^{-1} (p')\gamma_5 \frac {\tau^a }{2}
+ \gamma_5 \frac {\tau^a }{2} S^{-1} (p)
\nonumber
\]
for the axial vector vertex function
\begin{eqnarray*}
iS(p') &\Lambda_{\mu }^a &(p',p) iS(p )
\equiv \nonumber \\
&\int& d^4  x d^4 y e^{ip'x - ipy} \langle
TA_{\mu }^{a}(0) \psi (x) \bar \psi (y)\rangle
\nonumber
\end{eqnarray*}
and the quark propagator $S(p)$. The Ward identity holds in the chiral limit
and follows from the canonical commutation relations of the quark fields.
We investigate the propagator and the vertex for on-shell quarks,
with $p^2 =p'^2 =m^2$ and $S^{-1} = \overlay{\slash}{p} - m$.
For small momentum transfer $q=p'-p$ the vertex function is determined by
$g_A$, the pseudo scalar and pseudo vector pion quark coupling constants,
$g_{\pi q}^P $ and $g_{\pi q}^A $:
\[
\Lambda_{\mu }^a (p',p)
= \gamma_{\mu }  \gamma_5 \frac {\tau^a }{2} g_A
 - q_\mu \frac {F_\pi } {q^2}
    \left( g_{\pi q}^P -\frac {g_{\pi q}^A }{2m} \overlay{\slash}{q} \right)
  \gamma_5 \tau^a + \dots.
\nonumber
\]
The Goldberger-Treiman relations
$1-g_A =g_{\pi q}^A F_\pi /m $ and
$ g_{\pi q}^P F_{\pi} = m $, follow from evaluating the Ward
identity in the limit $q^2 = 0 $. We note that Ref. \cite{Vogl1,Vogl2}
derives these relations in the NJL model.
\bibitem{rem:infty}
{Note that in the approach of Ref.~\cite{Weinberg:nc} one has to use $g_A=1$ in
  the cross sections in the sum rule. Otherwise, if $g_A \neq 1$, the sum rule
  is divergent since the integration limit extends to infinity. However, at
  large values of $\omega$ the low-energy amplitude is not realistic. In this
  paper we introduce a cutoff for the low-energy process. Another possibility
  is to modify the amplitude. In the $\pi$-$N$ scattering one usually perform
  the $K$-matrix resummation in the $s$-channel \cite{Hoeler}. For the present
  case that would improve the convergence and ``save'' the sum rule. It would
  also modify the result numerically, but at an insignificant level.}
\end{references}
\end{document}